\documentstyle{article}
\title{\bf Protein Folding and cosmology}
\author{ {\bf Pedro F. Gonz\'alez-D\'{\i}az$^{*}$$^{\ddag}$}
and {\bf Carmen L. Sigenza$^{\dagger}$}\\
$^{*}$Instituto de Matem\'aticas y F\'{\i}sica Fundamental\\
Consejo Superior de Investigaciones Cientificas\\
Serrano 121, 28006 Madrid (SPAIN)\\
$^{\dagger}$ Universidad San Pablo CEU, Facultad de Ciencias Experimentales\\
y T'cnicas, Carretera de Boadilla del Monte, Km. 5.3,\\
28668 Boadilla del Monte (Madrid) (SPAIN)
}
\date{June 4, 1997}
\begin{document}
\maketitle
\large
\setlength{\baselineskip}{0.5cm}
%\vspace{3cm}

\renewcommand{\theequation}{\thesection . \arabic{equation} }

Protein denaturing induced by supercooling is interpreted as a
process where some or all internal symmetries of the native
protein are spontaneously broken. Hence, the
free-energy potential corresponding to a folding-funnel
landscape becomes temperature-dependent and describes a
phase transition.
The idea that deformed vortices could be produced in the transition
induced by temperature quenching, from native proteins to unfolded
conformations is discussed in terms of the Zurek mechanism that
implements the analogy between vortices, created in the laboratory
at low energy, and the cosmic strings which are thought to have been
left after symmetry breaking phase transitions in the early universe.
An experiment is proposed to test the above idea which generalizes the
cosmological analogy to also encompass biological systems and push a
step ahead the view that protein folding is a biological equivalent
of the big bang.

\vspace{4cm}

\noindent $^{\ddag}$ Corresponding author (Tel: (34-1) 5616800,
Fax: (34-1) 5645557, E-mail: iodpf21@cc.csic.es)

\pagebreak

\section{\bf Introduction}

The folding of proteins can be regarded as a biological equivalent
of the cosmological big bang in that its end result is
evident everywhere, in every living cell, but its beginnings and
main cause are shrouded in mistery. This is of course mere analogy,
but connections of the problem of protein folding with basic
aspects of fundamental physics are being discovered that call for
a more serious interrelation. In particular, many proteins at
sufficiently low temperature show a dynamical behaviour that
matches that of glasses and spin glasses, while offering a lot
to be earned about the physics of complexity (Frauenfelder \&
Wolynes, 1994).

In this paper, we push forward the above analogy in still another
important respect: we look at protein unfolding in supercooled
pure solvent as being a process where some internal symmetries
of the native protein are spontaneously broken. On the other
hand,
phase transitions induced by spontaneous symmetry breaking which
are thought to have occurred in the early universe could have left
behind long-lived topologically stable structures such as monopoles,
strings, domain walls and textures (Zel'dovich {\it et al.}, 1975;
Kibble, 1976; Albrecht \& Turok, 1985; Vilenkin \& Shellard, 1994).
The possible role played
by these structures in the generation of the present configuration
of the observable universe has been very much discussed in recent
decades (Zel'dovich, 1980; Vilenkin, 1981; Turok, 1983; Turok \&
Brandenberger, 1986).
Therefore, the use of superfluid helium (Hendry {\it et al.}, 1994;
B"uerle {\it et al.}, 1996; Ruutu {\it et al.}, 1996)
or liquid crystals (Chuang {\it et al.}, 1991; Bowick {\it et al.},
1994) to check in the laboratory the properties
of the cosmological defects has become a matter of great interest.
The possibility of performing such cosmological experiments in
helium was first suggested by Zurek (1985) who noted the analogy
between cosmological strings and vortex lines in the superfluid.
Here, this analogy is extended to also involve vortex lines filled
with folded conformations of proteins rapidly denatured by
lowering the temperature.

\section{\bf Protein folding and spontaneous symmetry breaking}
\setcounter{equation}{0}

Globular proteins which denature in water by lowering the temperature
can be described as undergoing a phase
transition driven by spontaneous breakdown of the internal
symmetries hidden inside the core of their folded, native structure.
Proteins in aquous medium interact with this through their hydrophobic
and hydrophilic (polar) groups. At the sufficiently high
temperatures for which no collective "coherent" effect can be
expected to occur in the whole system, each protein molecule
stabilizes by individually hiding most of its hydrophobic
groups, placing them in the interior of the native, compact
structure. The symmetries present in this structure can
therefore be characterized by the number of contacts among
its hydrophobic groups, and that number will tend to decrease
as the protein unfolds (Dill \& Su Chan, 1997).
At lower temperatures, water can still
act like a collective field that attracts protein molecules
to one another through the combined effect of the hydrophilic
groups and water-to-water hydrogen bonding.

The fundamental degrees of freedom here are the molecules themselves.
A way to characterize the state with the thermodynamically most
stable conformations is by the number density of protein molecules
in the different conformations
that exist, or that have "condensed" into this state. Since at sufficiently
high temperatures the unfolded protein molecules do not attractively
interact with each other (i.e. their mutual weak attractive
interaction is overcome by the combined effect of repulsive
interactions among water and hydrophobic groups and thermal energy),
the more unfolded conformations are formed the more energy
it takes. However, because protein molecules are much less
rigid in their unfolded conformations, water-mediated attractive
interaction among the hydrophilic groups of two or more
unfolded molecules is expected to be greater than that for
protein molecules in the native conformation. Therefore, when
the temperature becomes low enough to allow water molecules
to tend to grouping around the polar groups of the protein,
so that attractive interactions start dominating,
the total energy of the system
begins decreasing with the number density of unfolded conformations,
until the system saturates in them. At such temperatures, one would expect
all or some of the internal symmetries in the core of folded proteins
to be spontaneously broken, creating a more stable ensemble of
unfolded protein conformations.

Let us consider
the simplest form of a unidimensional section of
the effective free-energy potential contribution
satisfying the Anfinsen's requirement (Anfinsen, 1973)
that native structures
be thermodynamically stable states with conformation at the
global minimum of its accessible free energies
when the internal symmetry
is not spontaneously broken, so as the new view that
proteins change tertiary structure according to a folding funnel
(Dill \& Sun Chan, 1997)
of their accessible energy landscape. In order to mathematically
describe the phase change which is associated with protein
unfolding induced by supercooling the pure solvent, let us
introduce the quantity $\Psi$ -the order parameter- which
determines to what extend the geometrical distribution of
hydrophobic groups in the unfolded phase at low temperature
differs from that in the native phase, and adscribe
a value $\Psi=0$ to this (symmetric) phase and a nonzero value
of $\Psi$ to the (nonsymmetric) unfolded protein phase. The
order parameter $\Psi$ and the free-energy potential defined
in terms of $\Psi$ will respectively be associated with
a given normal mode or combination of normal modes
of the protein molecule, and with the corresponding section
of the folding funnel landscape.

Since the native protein is characterized by $\Psi=0$, near
the low-temperature transition point one can write the free-energy
potential as a power series in $\Psi$. The number of relevant
terms in this expansion should include those terms which are going to
allow denaturation of the protein also at high temperature; i.e.
we ought to cut off the expansion series only after the term
$\Psi^{8}$ to allow the existence of some local minima at
$\Psi\neq 0$ (visualizable at temperatures sufficiently higher
than the transition temperature), while ensuring $\Psi=0$ to
be the global minimum of the symmetric phase. Thus, a
simplest free-energy potential that satisfies these requirements
can be written in the form
\begin{equation}
V(\Psi)=\frac{1}{2}\alpha_1\Psi^2+\frac{1}{4}\alpha_2\Psi^4
-\frac{1}{6}\alpha_3\Psi^6+\frac{1}{8}\alpha_4\Psi^8,
\end{equation}
where the $\alpha_i$'s are
constants with the dimensions of a free energy, $i=1,...,4$, such
that $\alpha_1>\alpha_2>\alpha_3>\alpha_4$, and, since
the states with $\Psi=0$ and $\Psi\neq 0$ are assumed to possess
different symmetry, the coefficient for the linear term has
been neglected. Besides, we have restricted to the most
interesting case which is prepared to accommodate transitions
that follow a continuous process, rather than those occurring
at isolated points, and therefore all the odd-power terms
have been also neglected (otherwise, the transition points
would be fixed by setting to zero more than one equation
for the coefficients.)

Apart of the global minimum at $\Psi=0$, the potential free-energy
(2.1) possesses one local minimum and one local maximum. Spontaneous
symmetry breaking will occur in (2.1) when coefficient $\alpha_1$
is allowed to take on negative values. The potential
free-energy with coefficients $\alpha_{i}$ constant, resulting
from changing sign of $\alpha_{1}$ in (2.1)
will then show a local maximum at $\Psi=0$, two
local minima and one local maximum at $\Psi>0$. For the position of these
extrema to be at real values of $\Psi$ in the temperature-independent
approximation, the coefficients $\alpha_i$
should moreover satisfy
\begin{equation}
\alpha_1\left[\frac{1}{4}\alpha_1\alpha_4^2\pm\left(\frac{\alpha_3^2}{27}
-\alpha_2\alpha_4\right)\right]
+\frac{\alpha_2^2}{27}\left(\alpha_2\alpha_4-\frac{\alpha_3^2}{4}\right)\leq 0,
\end{equation}
where the lower signs correspond to potential (2.1) and the upper signs
to the potential with negative $\alpha_1$. Manipulating these expressions,
we obtain additional conditions on the $\alpha_i$'s:
\begin{equation}
\alpha_2^2>27\alpha_1\alpha_3, \; \; \alpha_3^2>\frac{27}{4}\alpha_2\alpha_4 .
\end{equation}

If we consider the dependence of the free-energy potential
on the thermodynamic quantities $P$, $T$ of the protein,
then the corresponding
thermodynamic potential should be represented as a function
$V(P,T,\Psi)$, where $\Psi$ must be determined from the
thermal equilibrium condition for which $V$ is a global minimum.
Restricting to processes taking place at constant pressure, it
follows that $\alpha_{1}(T)$ must vanish at the transition
point because this term must be larger than zero in the
symmetric phase (to have a global minimum at $\Psi=0$) and
smaller than zero in the phase with broken symmetry to
account for the existence of a global minimum at $\Psi\neq 0$.
Besides, for the transition point at which $\alpha_{1}$ vanishes
to be a stable state, it is necessary that the last term
$\alpha_{4}$ be positive and constant. However, one can
expect that the terms $\alpha_{2}(T)$ and $\alpha_{3}(T)$
still show a singularity where they change sign as function
of $T$. These singularities would correspond to new
transitions occurring at temperatures which must differ from
the main transition temperature. Thus, for a given $P=$const.,
one may write $\alpha_{1}\propto(T-T_{c1})$,
$\alpha_{2}\propto(T-T_{c2})$ and 
$\alpha_{3}\propto(T-T_{c3})$,
and, given the shape of the potential, $T_{c1}<T_{c2}<T_{c3}$.

Once the temperature-independent potential (2.1) is fixed, the
above results can also be obtained by using the following
procedure.
If we were dealing with a field theory in which coordinate $\Psi$
was taken to be the matter field, then we had a precise prescription
to investigate the symmetry behaviour of the effective theory with
negative $\alpha_1$ at nonzero temperature (Linde, 1979).
However, what we,
instead, have is a phenomenological theory with a potential linear
in free energy which is not but a generalization of
that of Ginzburg and Landau for
superconductivity (Ginzburg \& Landau, 1950).
Field theory, in turn, should be expected
to be nothing but a
covariant generalization of the Ginzburg-Landau theory for
a potential given in terms of energy
densities rather than thermodynamic potentials (free energies).
To obtain a temperature-dependent potential in our phenomenological
theory using the field-theoretical prescription (Linde, 1979),
we note that the potential of the Ginzburg-Landau theory
can be formally reproduced by regarding the order parameter as a
field, and using then the field-theoretical
prescription as applied to a "fermionic-like" field, rather
than the usual boson case. The ultimate reason to follow this formal
procedure resides in the fact that the Ginzburg-Landau
potential, or potential (2.1), is linear in free energy, rather
than energy-density.

Starting with the temperature-independent
potential (2.1), we shall extend this trick to attain the corresponding
temperature-dependent effective potential for our unidimensional
symmetric slice of a protein folding funnel landscape.
The Lagrange equation for the "field" coordinate $\Psi$ in the
broken symmetry phase would be given by
\begin{equation}
\left(\Box+\alpha_1-\alpha_2\Psi^2+\alpha_3\Psi^4-\alpha_4\Psi^6\right)\Psi=0.
\end{equation}
We then tentatively interpret the global minimum in (2.1) as
defining a "conformational vacuum". Therefore, in Eqn. (2.4)
we should first insert a shift $\Psi\rightarrow\bar{\Psi}
=\Psi+\sigma(T)$ to recover the picture where we can define
usual creation and annihilation operators with vanishing
conformational vacuum expectation for the "field" $\Psi$,
and then take the Gibbs average (Landau \& Lifshitz, 1975)
$\langle...\rangle=
\frac{{\rm Sp}\left[\exp\left(-\frac{H}{T}\right)...\right]}{{\rm Sp}\left[\exp\left(-\frac{H}{T}\right)\right]}$,
with $H$ the Hamiltonian and $T$ the temperature. We then obtain
\[\Box\sigma(T)+\alpha_1\sigma(T)
-\alpha_2\left(\sigma(T)^3+3\sigma(T)\langle\Psi^2\rangle\right)\]
\begin{equation}
+\alpha_3\left(\sigma(T)^5+10\sigma(T)^3\langle\Psi^2\rangle\right)
-\alpha_4\left(\sigma(T)^7+21\sigma(T)^5\langle\Psi^2\rangle\right)=0,
\end{equation}
where we have used the condition $\langle\Psi\rangle=0$ for the
new "conformational vacuum", and restricted to work in the lowest
order in $\alpha_2$, $\alpha_3$ and $\alpha_4$, where the quantities
$\langle\Psi^3\rangle$, $\langle\Psi^5\rangle$ and
$\langle\Psi^7\rangle$ vanish, and the contribution
from $\langle\Psi^4\rangle$ can be discarded.

Since in the harmonic approximation at constant $\sigma$,
$\alpha_1\langle\Psi^2\rangle$ is linear in the free-energy,
this quantity can now be calculated according to the fermionic
rules as
\begin{equation}
\alpha_1\langle\Psi^2\rangle
=\frac{1}{(2\pi)^3}\int\frac{d^3p}{2\omega_p^2}(2n_p-1),
\end{equation}
in which $\omega_p$ is the energy of the particles with
momentum $p$ and mass $m$, and $n_p$ is the occupation number.
Taking $n_p=\left(\exp\frac{\omega_p}{T}+1\right)^{-1}$,
discarding the temparature-independent term $-\int d^3p/2\omega_p^2$,
which can be eliminated by mass renormalization at $T=0$,
and recalling (Linde, 1979)
that all interesting effects should take place at
$T\gg m$, expression (2.6) becomes
\begin{equation}
\alpha_1\langle\Psi^2\rangle=
\frac{1}{2\pi^2}\int_{0}^{\infty}\frac{dp}{e^{\frac{p}{T}}+1}=\frac{T\ln 2}{2\pi^2},
\end{equation}
so that we would finally have a Lagrange equation that again corresponds to
an effective potential free-energy
given by
\begin{equation}
V=-\frac{1}{2}\alpha_1\epsilon_1\Psi^2+\frac{1}{4}\alpha_2\epsilon_2\Psi^4
-\frac{1}{6}\alpha_3\epsilon_3\Psi^6+\frac{1}{8}\alpha_4\Psi^8,
\end{equation}
where the relative temperature $\epsilon_j$ are defined as:
\begin{equation}
\epsilon_j=\frac{T_{cj}-T}{T_{cj}}, \;\; j=1,2,3
\end{equation}
with the critical temperatures
\begin{equation}
T_{c1}=f_1\frac{\alpha_1^2}{\alpha_2},\;\;
T_{c2}=f_2\frac{\alpha_1\alpha_2}{\alpha_3} > T_{c1},\;\;
T_{c3}=f_3\frac{\alpha_1\alpha_3}{\alpha_4} > T_{c2},
\end{equation}
in which conditions (2.3) have been used and the dimensionless numerical
coefficients $f_1$, $f_2$ and $f_3$ are all of order unity. The same
approximate result
would also be obtained if we took into account the discarded
terms proportional to $\langle\Psi^4\rangle$ in (2.5), since
such terms, which can be calculated
by the above procedure,
would contribute $T_{c1}$ and $T_{c2}$ with
a factor of order unity because
of conditions (2.3). These results can readily be generalized to the
more realistic case of a potential with an arbitrary number of
higher-order potential terms, the highest of which being even
and positive, and terms with odd power of $\Psi$ which break
invariance under $\Psi\rightarrow-\Psi$. For the purposes of
this work, however, it will suffice working with the simplest
potential slice (2.8).

Note that (2.8) reduces in fact to the typical thermodynamical
potential of the Ginzburg-Landau
theory (Ginzburg \& Landau, 1950)
in the limit $\alpha_3, \alpha_4\rightarrow 0$, with
$\alpha_4$ going to zero more rapidly than $\alpha_3$.
The considered
phase transition for protein folding can
therefore be regarded as a generalization from
the second-order phase transitions taking place in the
phenomenological theory of superconductivity. At
$T_{c1}$ there will be a second-order phase transition from a potential
with a shape as in (2.1) to the corresponding potential with the
broken internal symmetries.
According to (2.8), other transitions can also occur at the generally
higher critical temperatures $T_{c2}$ and $T_{c3}$, these taking place between
states with the folded protein at the global minimum.

\section{\bf Cosmology in a protein}
\setcounter{equation}{0}

Like in superconductors and superfluids (Tilley \& Tilley, 1986),
or field theories (Nielsen \& Olesen, 1973),
in the present phenomenological model strings filled with a protein
in its symmetric, native phase will be able to form
when the order parameter $\Psi$ is allowed to be complex
\begin{equation}
\Psi=|\Psi|e^{i\theta},
\end{equation}
where $\theta$ is a fixed phase. The possibility of formation of
vortex lines in the considered phase transition in proteins
required therefore having a two-dimensional section of the
inverted folding funnel (Dill \& Su Chan, 1997)
that corresponded to the
breakdown of a complete given symmetry. Folded proteins contain
a lot of internal symmetries which can be visualized as contacts
between hydrophobic groups in the protein core
(Dill, 1995; Dill \& Su Chan, 1997). Each of
these symmetries can be independently broken by variation of
one, two or possibly more normal coordinates. In the first
case, a domain wall could be formed; when variation of two
normal coordinates were required to break down the hydrophobic
contacts defining one of such internal symmetries, then one
would expect the formation of stringy topological defects, a
case which we assume to hold hereafter.

Generally, strings can be formed because in the broken-symmetry
phase where $T<T_{c1}$, the first term in (2.1) is negative and,
since the last term of this potential is always positive, $V$
has the shape of a wrinkled mexican sombrero; then, the
manifold of the most stable conformations (conformational
"vacuum" manifold) is not simply connected and, therefore,
possesses nontrivial loops. By the first homotopy group
(Kibble, 1980; Vilenkin \& Shellard, 1994)
it follows that axisymmetric topological defects (vortex lines)
should then be expected to form along the phase transition at
temperature $T_{c1}$. This can be most clearly understood by
looking at the broken-symmetry phase with inverted folding
funnel potential as a quantum system.

Below the critical temperature $T_{c1}$, a temperature-dependent
fraction of protein macromolecules become trapped in their
various unfolded conformations and, together with the solvent
molecules, will form a phase which can be described by a wave
function $\Psi(r)$. The properties of the protein-solvent complex
in the broken-symmetry phase can be understood assuming that this
wave function satisfies a wave equation of the form
\begin{equation}
i\hbar\frac{\partial\Psi}{\partial t}=-\frac{\hbar^2}{2m}\nabla^2\Psi
+\mu\Psi ,
\end{equation}
where $m$ is taken to be the mass of the complex formed by a protein
macromolecule and its hydration shell, and $\mu$ is the chemical
potential, i.e. the energy gained by the system when one molecule
of protein is unfolded at constant volume and entropy.

Let us assume that the free energy can be expanded in powers of
$|\Psi|^2$ (which would now play the role of the order parameter)
and has the form:
\begin{equation}
F(r)=\alpha_1(T)|\Psi|^2+\frac{1}{2}\alpha_2(T)|\Psi|^4
-\frac{1}{3}\alpha_3(T)|\Psi|^6+\frac{1}{4}\alpha_4|\Psi|^8 ,
\end{equation}
where $\alpha_j(T)=\epsilon_j\alpha_j$.

The chemical potential $\mu$ can be evaluated by using (3.3) so
that the Schr"dinger equation becomes
\begin{equation}
i\hbar\frac{\partial\Psi}{\partial t}=-\frac{\hbar^2}{2m}\nabla^2\Psi
+\alpha_1(T)\Psi+\alpha_2(T)|\Psi|^2\Psi
-\alpha_3(T)|\Psi|^4\Psi+\alpha_4|\Psi|^6\Psi.
\end{equation}
Eq. (3.4) can now be re-scaled to assure a dimensionless form, which is
given by
\begin{equation}
i\dot{\eta}=-\nabla^2\eta+\left(-1+|\eta|^2-|\eta|^4+|\eta|^6\right)\eta ,
\end{equation}
by introducing the quantities
\begin{equation}
\tau_1=\frac{\hbar}{|\alpha_1(T)|}=\frac{\tau_{01}}{|\epsilon_1|},\;\;
\tau_{01}=\frac{\hbar}{\alpha_1}
\end{equation}
\begin{equation}
\xi_1=\frac{\hbar}{\sqrt{2m|\alpha_1(T)|}}=\frac{\xi_0}{\sqrt{\epsilon_1}},
\;\; \xi_0=\sqrt{\frac{\hbar\tau_{01}}{2m}}
\end{equation}
\begin{equation}
\sigma_j^2=-\frac{\alpha_1(T)}{\alpha_{j\neq 1}(T)}
=-\frac{\alpha_1\epsilon_1}{\alpha_{j\neq 1}(T)},
\end{equation}
where $\tau_1$ is to be interpreted as the relaxation time,
$\xi_1$ as the correlation length, and the $\sigma_{j}^2$'s as
equilibrium linear densities characterizing the broken-symmetry
phase for the unfolded protein. In these definitions, $\tau_{01}$
is a time parameter that also characterizes the phase with broken
symmetry.

When we restrict ourselves to deal with the time-independent
solutions to Eq. (3.5)
in cylindrical coordinates ($\rho$,$\phi$,$z$), i.e.
\begin{equation}
\frac{\partial\eta}{\rho\partial\rho}+\frac{\partial^2\eta}{\partial\rho^2}
+\frac{\partial^2\eta}{\rho^2\partial\phi^2}=
\left(|\eta|^2-|\eta|^4+|\eta|^6-1\right)\eta ,
\end{equation}
one has the trivial solution $|\eta|^2=1$ and the axisymmetric solution
\begin{equation}
\eta=f(\rho)\exp(in\phi),
\end{equation}
with $n$ a whole number. The radial part of (3.10) is regular for
$\rho\ll 1$, where $f(\rho)\simeq\rho^n$, and approaches the
equilibrium value of the trivial solution at $\rho\gg 1$, for
which $f^2(\rho)\simeq 1-n^2/2\rho^2$. Thus, solution (3.10)
represents a stringy topological defect which, for $n=1$,
is still of the
kind first considered by Ginzburg and Pitaevskii in the context
of the phenomenological theory of superfluidity
(Ginzburg \& Pitaevskii, 1958). Around the
axis of symmetry, there will be a flow caused by the phase gradient,
with velocity $v=\hbar\nabla\theta/m=\hbar/mr$. The resulting
vortex will be filled with protein molecules keeping the native
structure, or some partially folded conformations, and will be
surrounded by a distribution of distinct unfolded configurations.
It appear that a certain nonzero density of these vortex lines
would be formed if the second-order phase transition is induced
by a sufficiently rapid temperature quenching.

However, the isotropy assumption for the order parameter, by
which variation of the accessible free energy respect to normal
coordinates $\Psi_1$ and $\Psi_2$ are regarded to be equal,
is obviously not but an approximation. Strictly speaking, the
cylindric coordinates cannot be circular, but only closed on
the $z$-constant sections. This would lead to stringy solutions
corresponding to a cylinder deformed on its $z$-constant
sections and closed flow streamlines no longer perfectly circular.
These deformed solutions would satisfy (3.10) only approximately.

We shall estimate the density of vortex lines formed during the
phase transition of a protein induced on a dynamical time scale
by a rapid decrease of temperature. In the vecinity of $T_{c1}$,
the relaxation time scale $\tau_1$ (i.e. the time scale on which
the order parameter can adjust to the new thermodynamic parameters)
will become quite longer than the time on which the quench proceeds.
Then, the order parameter $\Psi$ will be practically frozen on that
time scale and correlated over large domains where $\xi_1\rightarrow\infty$
({\it impulse} regime). Sufficiently far from $T_{c1}$, $\tau_1$
will be much smaller that the quench time and $\Psi$ will become
at an equilibrium configuration with $\xi_1$ determined by the
instantaneous value of $\epsilon_1$ ({\it adiabatic} regime)
(Zurek, 1996).

If the transition is made fast enough, then the growing regions
of the new phase met and coalesced, making it impossible to
avoid the creation of topological defects. That will happen
on the boundary between the impulse and adiabatic regimes,
at the freeze out time $\hat{t}$ (Zurek, 1996), if
\begin{equation}
R_b >\hat{\xi_1},
\end{equation}
where $R_b$ is the effective radius of the region with the starting
symmetric phase, and $\hat{\xi_{1}}$ is the characteristic correlation
length at the freeze out time $\hat{t}$.

At a sufficiently rapid pace, on a quench time scale $\tau_Q$
(controlled by the rate at which the temperature is lowered),
one can assume (Zurek, 1996)
that $\epsilon_1$ is proportional to time,
$\epsilon_1=\frac{t}{\tau_Q}$,
in the vecinity of $T_{c1}$. Thus, at $\hat{t}$ there will be a
freeze out temperature $\hat{\epsilon_1}=\hat{t}/\tau_Q$, so that
using (3.6), we obtain
$\hat{t}=\sqrt{\tau_{01}\tau_Q}$.
The transition between the adiabatic and impulse regimes will
therefore take place at the relative temperature
$\hat{\epsilon_1}=\epsilon(\hat{t})=\sqrt{\tau_{01}/\tau_Q}$,
corresponding to a freeze out correlation length which, using
(3.7), can be expressed as
\begin{equation}
\hat{\xi_1}=\frac{\xi_0}{\sqrt{\hat{\epsilon_1}}}
=\sqrt{\frac{\hbar}{2m\alpha_1}}(\tau_{01}\tau_Q)^{\frac{1}{4}}
=\left(\frac{\hbar^3\tau_Q}{4m^2\alpha_1}\right)^{\frac{1}{4}}.
\end{equation}

Since $\alpha_1$ corresponds to an free energy which may be
associated with a large number of the protein internal motions,
such as unfolding of $\alpha$-helices
(Creighton, 1992) and hindered rotations,
each with energies typically of the order $10^{-13}$ erg., for most
proteins $\alpha_{1}$ will take on values within the interval
$(10^{-10} - 10^{-13})$ erg., and hence $\hat{\xi_{1}}$ will have
values in the interval
$(10^{-6} - 10^{-7})\tau_Q^{\frac{1}{4}}$ cm. For sufficiently rapid,
but still
not very fast temperature quench, we then obtain $\hat{\xi_{1}}\ll 1$
cm, and therefore there will be a copious initial production of
vortex lines with density $\hat{\xi_1}^{-2}$ (Zurek, 1996).

We propose finally an experiment to check the possible formation
of vortex lines and other topological defects during the phase
transition leading from folded to unfolded conformations in
apomyoglobin (aMb). The transition will be induced by rapid
supercooling, following excitation by focused pulses of a tunable
infrared laser. The experiment is a variant of that recently
carried out by Ballew {\it et al.} (1996) in order to
follow the main fastest events of the aMb folding. It would
consist of first denaturing aMb by supercooling it in water
or D$_2$O at around -10 C in a perfectly insulated
cell equipped with a high
precision device to monitor variations of energy locally.
This could be e.g. made by measuring the transmission
of a calibrated low-power diode laser focused at the given
region. The sample would be held in a cuvette with optical
path length of the order 1 mm, cooled by thermoelectric devices
keeping the temperature constant at around -10 C, or inducing
a fast recovery of that temperature, in the region previously
heated by the infrared laser, with a thermistor feedback loop.
Two counterpropagating beams from an infrared laser would here
coherently and uniformly heat a cylindric region with section
$\sim$1 mm$^{2}$, of the previously cooled sample, inducing the
folding of aMb in the resulting irradiated volume. The pulsed infrared
laser delivered short pulses ($\sim$ 1 ns) at wavelengths able
to excite a near-infrared mode of water only
(Ballew {\it et al.}, 1996). After absorbing
the infrared photons, water will almost instantly transform all
absorbed energy into heat that locally induced naturation of aMb.

After the laser excitation, the created interphase would inmediately
start traveling inward the irradiated volume, quenching it to
the denatured conformations in a time $\tau_Q$ short enough
to satisfy (3.11), so creating a tangle of topological defects
filled with folded or partly folded aMb conformations. The
presence of these defects would finally be detected by
measuring the deficit in energy released as, in the
irradiated volume, aMb molecules in the symmetric phase pass back
into the low-temperature unfolded phase. By ascribing this
deficit to the formation of vortices and other topological
defects (B"uerle {\it et al.}, 1996),
one could infer the resulting defect density.

\vspace{1.1cm}

\noindent {\bf Acknowledgements}.
P.F.G.-D. acknowledges DGICYT by support under Research Project
No. PB94-0107.

\pagebreak
\noindent\section*{References}
\begin{description}
\item [] ALBRECHT, A. \& TUROK, N. (1985). Evolution of cosmic strings.
{\it Phys. Rev. Lett.} 54, 1868-1871.
\item [] ANFINSEN, C.B. (1973). Principles that govern the folding
of protein chains.
{\it Science} 181, 223-230.
\item [] BALLEW, R.M., SABELKO, J. \& GRUEBELE, M. (1996).
Direct observation of fast protein folding: The initial collapse
of apomyoglobin. {\it Proc. Natl. Acad.
Sci. USA} 93, 5759-5764.
\item [] BAEUERLE, C., BUNKOV, Yu. M., FISHER, S.N., GODFRIN, H.
\& PICKETT, G.R. (1996). Laboratory simulation of cosmic string formation
in the early universe using superfluid $^{3}$He. {\it Nature} 382, 332-334.
\item [] BOWICK, M.J., CHANDAR, L., SCHIFF, E.A. \& SRIVASTAVA, A.M. (1994).
The cosmological Kibble mechanism in the laboratory: String formation
in liquid crystals.
{\it Science} 263, 943-945.
\item [] CHUANG, I., DUERRER, R., TUROK, N. \& YURKE, B. (1991).
Cosmology in the laboratory: Defect dynamics in liquid crystals.
{\it Science}
251, 1336-1342.
\item [] CREIGHTON, T.E. (1992). {\it Protein Folding} (W.H. Freeman and Co.,
New York, USA).
\item [] DILL, K.A. {\it et al}. (1995). Principles of protein foldings
-A perspective from simple exact models.
{\it Protein Sci.} 4, 561-602.
\item [] DILL, K.A. \& SUN CHAN, H. (1997).
From Levinthal to pathways to funnels.
{\it Nature Struct. Biol.} 4, 10-19.
\item [] FRAUENFELDER H. \& WOLYNES, P.G. (1994).
Biomolecules: Where the physics of complexity and simplicity meet.
{\it Phys. Today}, February, 58-64.
\item [] GINZBURG, V.L. \& LANDAU, L.D. (1950).
On the theory of superconductivity.
{\it Zh. 'ksp. teor. Fiz.} 20, 1064-1082.
\item [] GINZBURG, V.L. \& PITAEVSKII, L.P. (1958).
On the theory of superfluidity.
{\it Zh. 'ksp. teor. Fiz.}
34, 1240-1245; {\it Soviet Phys. JETP} 34, 858-863.
\item [] HENDRY, P.C., LAWSON, N.S., LEE, R.A.M., McCLINTOCK, P.V.E.
\& WILLIAMS, C.H.D. (1994).
Generation of defects in superfluid $^{4}$He as an analogue
of the formation of cosmic strings.
{\it Nature} 386, 315-317.
\item [] KIBBLE, T.W.B. (1976). Topology of cosmic domains and strings.
{\it J. Phys.} A9, 1387-1398.
\item [] KIBBLE, T.W.B. (1980). Some implications of a cosmological
phase transition.
{\it Phys. Rep.} 67, 183-199.
\item [] LANDAU, L.D. \& LIFSHITZ, E.M. (1975).
{\it F¡sica Estad¡stica}
(Revert', S.A., Barcelona).
\item [] LINDE, A.D. (1979).
Phase transitions in gauge theories and cosmology.
{\it Rep. Prog. Phys.} 42, 389-437.
\item [] NIELSEN, H.B. \& OLESEN, P. (1973).
Vortex-line models for dual strings.
{\it Nucl. Phys.} B61, 45-61.
\item [] RUUTU, V.M.H., ELTSOV, V.B., GILL, A.J., KIBBLE, T.W.B.,
KRUSIUS, M., MAKHLIN, Yu. G., PLA€AIS, B., VOLOVIK, G.E. \&
SU, W. (1996).
Vortex formation in neutron-irradiated superfluid $^{3}$He as an
analogue of the formation of cosmic strings.
{\it Nature} 382, 334-336.
\item [] TILLEY, D.R. \& TILLEY, J. (1986). {\it Superfluidity and
Superconductivity}, 2nd edn. (Hilger, Boston, USA).
\item [] TUROK, N. (1983).
The evolution of cosmic density perturbations around grand unified strings.
{\it Phys. Lett.} 126B, 437-440.
\item [] TUROK, N. \& BRANDENBERGER, R.H. (1986).
Cosmic strings and the formation of galaxies and cluster of galaxies.
{\it Phys. Rev.} D33, 2175-2181.
\item [] VILENKIN, A. (1981).
Cosmological density fluctuations produced by vacuum strings.
{\it Phys. Rev. Lett.} 46, 1169-1172; Erratum:
{\it Phys. Rev. Lett.} 46, 1496.
\item [] VILENKIN, A. \& SHELLARD, E.P.S. (1994)
{\it Cosmic Strings and
other Topological Defects} (Cambridge Univ. Press, Cambridge, UK).
\item [] ZEL'DOVICH, Ya. B. (1980).
Cosmological fluctuations produced near a singularity.
{\it Month. Not. Roy. Astron. Soc.} 192, 663-667.
\item [] ZEL'DOVICH, Ya. B., KOBZAREV, I.Yu. \& OKUN, L.B. (1974).
Cosmological consequences of the spontaneous breakdown of
a discrete symmetry.
{\it Zh. 'ksp. teor. Fiz.} 67, 3-11; (1975) {\it Soviet Phys.} JETP 67, 401-409.
\item [] ZUREK, W.H. (1985).
Cosmological experiments in superfluid helium?.
{\it Nature} 317, 505-508.
\item [] ZUREK, W.H. (1996).
Cosmological experiments in condensed matter systems.
{\it Phys. Rep.} 276, 177-221.

\end{description}

\end{document}